\documentclass[pdflatex,sn-mathphys-num]{sn-jnl}

\usepackage{graphicx}%
\usepackage{multirow}%
\usepackage{amsmath,amssymb,amsfonts}%
\usepackage{amsthm}%
\usepackage{mathrsfs}%
\usepackage[title]{appendix}%
\usepackage{xcolor}%
\usepackage[normalem]{ulem}%
\usepackage{textcomp}%
\usepackage{manyfoot}%
\usepackage{booktabs}%
\usepackage{algorithm}%
\usepackage{algorithmicx}%
\usepackage{algpseudocode}%
\usepackage{listings}%

\newcommand{\be}{\begin{eqnarray}}
\newcommand{\ee}{\end{eqnarray}}

\newcommand{\ave}[1]{\left\langle #1 \right\rangle}

\newcommand{\mev}{{\rm \, MeV}}
\newcommand{\gev}{{\rm \, GeV}}
\newcommand{\fc}[1]{\hat{C}_#1}
\newcommand{\cpl}[1]{\hat{c}_#1}
\newcommand{\sNN}{\sqrt{s_{\rm NN}}}

\begin{document}

\title[Exploring the QCD phase diagram through correlations and fluctuations]{Exploring the QCD phase diagram through correlations and fluctuations}

\author*[1]{\fnm{Volker} \sur{Koch}}\email{vkoch@lbl.gov}

\author[2]{\fnm{Volodymyr} \sur{Vovchenko}}\email{vvovchen@central.uh.edu}
\equalcont{These authors contributed equally to this work.}

\affil*[1]{\orgdiv{Nuclear Science Division}, \orgname{Lawrence Berkeley National Laboratory}, \orgaddress{\street{1 Cyclotron Rd}, \city{Berkeley}, \postcode{94720}, \state{CA}, \country{USA}}}

\affil[2]{\orgdiv{Physics Department}, \orgname{University of Houston}, \orgaddress{\street{3507 Cullen Blvd}, \city{Houston}, \postcode{77204}, \state{TX}, \country{USA}}}

\abstract{The exploration of the Quantum Chromodynamics (QCD) phase diagram is a central goal of relativistic heavy-ion collision experiments. This review focuses on the role of fluctuations and correlations as sensitive probes of the phase structure. We discuss theoretical advancements and experimental methodologies employed to map the QCD phase diagram, highlighting constraints derived from both lattice QCD calculations and existing experimental data. Key observables such as cumulants and factorial cumulants of conserved charges (e.g., net-proton, net-charge) are explored as promising signatures of phase transitions and the QCD critical point. We discuss how these quantities are measured experimentally and compared with theoretical predictions, addressing challenges and best practices for meaningful comparisons. Special attention is given to predictions and current experimental results at high baryon density, including recent findings from the STAR collaboration at RHIC.  Finally, we identify open issues and future directions for fluctuation and correlation studies at lower collision energies, relevant for future measurements, for example by the CBM experiment.}

\maketitle

\section{Introduction}\label{sec:introduction}
One of the major goals in the physics of the strong interaction is the study of the
properties of strongly interacting matter, in particular its phase structure. This phase structure
is usually depicted in the QCD phase diagram, which is typically discussed in terms of the
temperature $T$ and the baryon chemical potential $\mu_B$.
Theoretically, the phase diagram is studied using thermal field theory,  most prominently lattice
QCD calculations, which provide a non-perturbative framework for understanding the behavior of QCD
at finite temperature and density. Lattice QCD calculations have shown that the QCD phase diagram features a crossover transition from hadronic matter to a quark-gluon plasma (QGP) at high temperatures and vanishing baryon chemical potential \cite{Aoki:2006we}. Unfortunately, due to the fermion sign problem, lattice QCD calculations are not feasible at high baryon chemical potential, which is the region of interest for the search of the QCD critical point (CP) and where many model calculations predict a first-order phase transition (see e.g. \cite{Stephanov:2004wx}). 
However, as we shall discuss, recently some progress has been made to constrain the position of the QCD critical point, either through extrapolations from lattice QCD calculations at vanishing (and imaginary~\cite{deForcrand:2002hgr,DElia:2002tig}) chemical potential \cite{Dimopoulos:2021vrk,Basar:2023nkp,Clarke:2024ugt,Shah:2024img}, or by applying functional methods such as Dyson-Schwinger equations \cite{Fischer:2014mda,Gao:2020fbl,Bernhardt:2021iql}, the functional renormalization group \cite{Fu:2019hdw,Fu:2023lcm}, as well as effective-model-based extrapolations constrained to lattice data~\cite{Hippert:2023bel}. Interestingly, most of these calculations predict a QCD critical point in the same region with a critical temperature of $100 \mev \lesssim T_C \lesssim 120 \mev$ and a critical baryon number chemical potential of $550 \mev \lesssim \mu_C \lesssim 650 \mev$. 
Using the standard freeze-out systematics this region would roughly correspond to a collision energy of $\sNN \simeq 5 \gev$. 

Experimentally, the QCD phase diagram is explored through high-energy heavy-ion collisions, such as
those conducted at the Large Hadron Collider (LHC) at CERN and the Relativistic Heavy Ion Collider
(RHIC) at Brookhaven National Laboratory. In these collisions, strongly interacting matter at high
temperature is created and hadrons in the central rapidity region are measured. 
Since the baryon number is conserved, the total net baryon number corresponds to the number of participant nucleons, i.e. no additional net-baryons will be created in these
collisions. 
At the highest energies, the net-baryon density at mid-rapidity is close to zero. 
Therefore, in order to create matter at finite net-baryon density one needs to lower the
beam energy so that the baryons from projectile and target nuclei stop in the mid-rapidity region
where measurements are typically carried out. 
Since the location of the QCD critical point is not known, the beam energy is varied in order to explore as large a region of the QCD phase diagram as possible. 
This was the motivation for the beam energy scan (BES) program at RHIC which has been designed to systematically explore the QCD phase diagram at finite baryon density (for a review of the results from the first phase of this program see \cite{Bzdak:2019pkr}). At lower energies, fixed-target experiments such as NA61/SHINE~\cite{NA61SHINE:2024xdd} at CERN and HADES at GSI~\cite{HADES:2020wpc} provide complementary fluctuation measurements that extend the coverage of the phase diagram to much higher baryon densities.

One key observable in the search for the QCD critical point is the fluctuations of conserved
charges, in particular baryon number fluctuations. The fluctuations are typically characterized by
the (factorial) cumulants of the baryon number distribution. As shown in \cite{Stephanov:2008qz} the
cumulants scale with powers of the correlation length which at the QCD critical point diverges, and the
higher the order of the cumulant the stronger the divergence. However, cumulants are not only useful
for identifying the QCD critical point. As we shall explain below, baryon number cumulants represent the
derivatives of the pressure (or grand potential) with respect to the baryon number chemical potential, $\mu_B$. 
Therefore, they are sensitive to any non-trivial structures in the pressure or free energy,
such as the cross-over transition of QCD at small chemical potential.
Being derivatives of the pressure with respect to the chemical potential, cumulants can also be calculated in thermal field theory, in particular lattice QCD albeit only at vanishing chemical potential. In theory, this allows for a systematic comparison of experimental results with theoretical predictions, although, as we shall discuss, such a comparison requires care. 

Another possibility to see, albeit indirect, signs for a QCD critical point, is by finding evidence
for the associated first-order phase coexistence region. 
This could be achieved by a suitable choice
of collision energy such that the system spends sufficient time in the mechanically unstable
spinodal region. 
The spinodal instability will lead to rapid phase separation producing lumps of
hadronic matter of a characteristic size \cite{Randrup:2003mu,Randrup:2009gp}. 
Spinodal clumping has been
successfully utilized to find evidence for the first-order liquid-gas transition of nuclear matter
\cite{Chomaz:2003dz}. That spinodal clumping should also happen during  the transition from hadronic matter to the quark-gluon
matter has been convincingly demonstrated in an explicit hydrodynamic calculation
\cite{Steinheimer:2012gc}. However, while the clumping is clearly visible in configuration space,
attempts to find measurable observables in momentum space, where experiments measure, have so far
failed \cite{Steinheimer:2013xxa,Steinheimer:2019iso}. 
This may be due to the lack of sufficient collective flow at the energies where the instability occurs, which translates the spatial correlations into measurable momentum correlations.
Another avenue for pursuing signatures of a first-order transition is through electromagnetic probes, such as dileptons, whose spectrum can be affected by the system spending time in the mixed phase region~\cite{Rapp:2014hha,Seck:2020qbx,Savchuk:2022aev}.

The CBM experiment at FAIR will study Au-Au collisions in the energy range $\sqrt{s_{\rm NN}} = 2.8 – 4.9\gev$, probing the region of the phase diagram where the aforementioned predictions seem to converge as to the possible location of the QCD critical point.
Utilizing high-luminosity beams and state-of-the-art detector capabilities, CBM is well-positioned to deliver the necessary precision measurements of high-order fluctuations and correlations, and related observables in the search for the QCD critical point.

Throughout this review we focus on fluctuation and correlation observables in relativistic heavy-ion collisions, while the constraints from lattice QCD and other theoretical methods are used as guidance for the region of the phase diagram accessible in experiments.

This review is organized as follows:  In Sec. \ref{sec:cp} we review the current status of the theoretical understanding of the QCD critical point. In Sec. \ref{sec:fluctuations} we discuss the role of fluctuations and correlations in the search for the QCD critical point, including experimental methodologies and challenges. In Sec. \ref{sec:exp_vs_theory} we present a comparison of experimental data with theoretical predictions, focusing on recent results from RHIC and LHC. In Sec.~\ref{sec:data_and_predictions} we discuss non-critical baselines and compare them with experimental results. In Sec.~\ref{sec:open_issues} we summarize the main lessons, discuss open issues and next steps, and outline future directions for research in this area. Finally, Sec.~\ref{sec:conclusions} provides an overall summary and outlook.

\section{Status of theoretical predictions for the QCD critical point}
\label{sec:cp}

\begin{figure}[h]
\centering
\includegraphics[width=.95\textwidth]{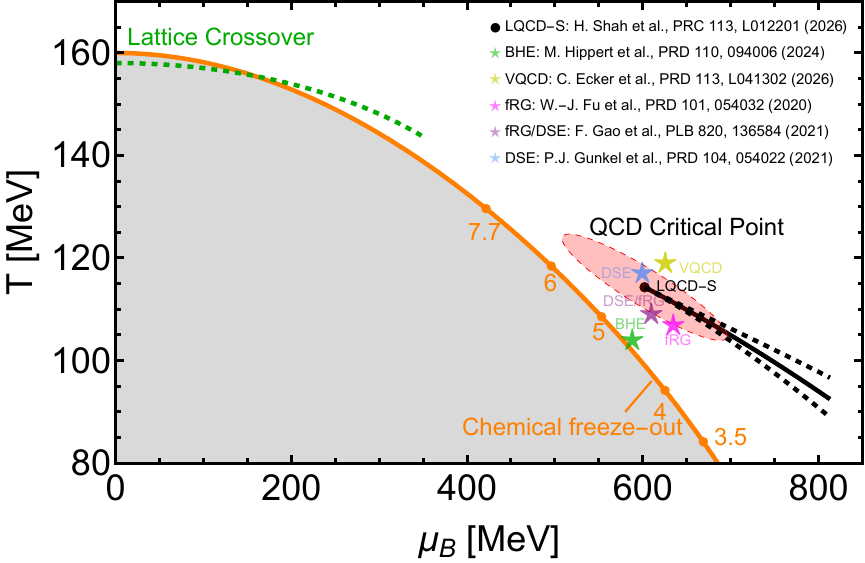}
\caption{Based on~\cite{Shah:2024img}.
A compilation of predictions for the location of the QCD critical point on the $T$-$\mu_B$ phase diagram of QCD. 
The black point with a red covariance ellipse shows the estimate from Ref.~\cite{Shah:2024img}, based on the extrapolation of constant entropy density contours from $\mu_B = 0$. 
The stars depict estimates from other approaches, functional methods~(fRG~\cite{Fu:2019hdw}, DSE-fRG~\cite{Gao:2020fbl}, DSE~\cite{Gunkel:2021oya}) and holography~(BHE~\cite{Hippert:2023bel}, VQCD~\cite{Ecker:2025vnb}).
The orange line represents the chemical freeze-out estimate from Ref.~\cite{Lysenko:2024hqp}, with points on the line corresponding to various collision energies~(in terms of $\sqrt{s}_{\rm NN}$ in GeV).
Both the chemical freeze-out line and all CP estimates~[except DSE ($\mu_s = 0$) and VQCD ($\beta$-equilibrium)] correspond to $\mu_S = 0$~($n_S \neq 0$) conditions.
The dashed green line depicts the chiral crossover line from~\cite{Borsanyi:2020fev}.
}
\label{fig:CPplot}
\end{figure}

As mentioned in the Introduction, lattice QCD calculations are not feasible in the region of interest for the search of the QCD critical point (CP) due to the sign problem at finite baryon density.
For this reason, predictions on the CP location must rely on extrapolations of lattice QCD calculations from vanishing baryon density, or on other methods.
In lattice QCD it is possible to calculate the derivatives of the pressure with respect to the chemical potential, the so-called baryon-number susceptibilities, for vanishing baryon number chemical potential \cite{Borsanyi:2012cr,Bellwied:2015lba,Bazavov:2017dus,HotQCD:2017qwq,Bazavov:2020bjn}.  
Alternatively, calculations at imaginary chemical potential are possible as they do not suffer from the fermion sign problem \cite{Borsanyi:2018grb}. Upon analytic continuation to real values of the chemical potential, these provide an alternative method to calculate these susceptibilities. 
The availability of higher order susceptibilities allows for a Taylor expansion of the pressure and lower order susceptibilities at small but finite chemical potential, $\mu_B/T \lesssim 3$. 
The application of generalized expansion schemes may allow one to extend the accuracy range of such expansions to somewhat higher net-baryon densities~\cite{Borsanyi:2021sxv,Kahangirwe:2024xyl,Abuali:2025tbd}.
Alternatively, one can also utilize the cluster expansion in fugacity space \cite{Vovchenko:2017gkg,Bellwied:2021nrt} with comparable range of validity in $\mu_B/T$.
However, such expansions by construction cannot incorporate a description of the CP and thus the validity range of these expansions necessarily falls short of where the CP may be.

A different strategy lies in isolating observables which may be uniquely sensitive to the CP. 
One such strategy is to use the Yang-Lee edge singularities~\cite{Skokov:2024fac} in the complex chemical potential plane.
These singularities are connected to the CP where they pinch the real axis~\cite{Stephanov:2006dn} and can thus be used to constrain the CP location.
Significant effort has been put into this in the lattice QCD community recently~\cite{Basar:2023nkp,Clarke:2024ugt,Adam:2025phc}.
First, one uses Pad\'e-type approximants from zero and imaginary chemical potential lattice calculations to determine complex chemical potential plane singularities that may be associated with the Yang-Lee edge singularities.
Then, one follows the trajectories of these singularities and extrapolates them to smaller temperatures. 
If the Yang-Lee singularities approach the real axis, this may signal the presence of the CP.
Current analyses suggest a possibility of the CP at a temperature $T \lesssim 110$ MeV~\cite{Basar:2023nkp,Clarke:2024ugt,Adam:2025phc}, below the temperatures where lattice QCD calculations have been performed.
The analysis, however, relies both on the validity of the Pad\'e approximants to determine Yang-Lee edge singularities and on accurate extrapolations to smaller temperatures. Furthermore, continuum extrapolation has not been achieved yet and, in addition, the lattice QCD results used are for finite volumes, i.e. no extrapolations to infinite volume have been done.
These systematic effects may significantly affect the corresponding predictions for the CP location, as also noted in Ref.~\cite{Clarke:2024ugt}. For this reason, the Yang-Lee edge singularity predictions are not included in the compilation plot Fig.~\ref{fig:CPplot}, although this promising approach may yield reliable estimates once these cutoff effects are under control.

A different extrapolation strategy has been put forward recently in~\cite{Shah:2024img} by using the constant entropy density contours.
This approach is rooted in the expected crossings of the constant entropy density contours at the CP, reflecting the associated singularity in the EoS.
The method is based on expansions along the constant entropy density contours, which involves both the susceptibilities and their temperature derivatives. Because the expansion is implicit in $T$, it permits a multi-valued behavior of the observables such as entropy density and leads to a mean-field type description of a first-order transition region.
This approach predicts a CP at $T \approx 114$ MeV and $\mu_B \approx 602$ MeV~\cite{Shah:2024img} under $\mu_S = \mu_Q = 0$ conditions~(see Fig.~\ref{fig:CPplot}).
Note that since the method is based on an expansion from $\mu_B = 0$, it can only capture mean-field critical behavior rather than the correct 3D Ising universality class expected for the QCD critical point.
In Ref.~\cite{Borsanyi:2025dyp} the Budapest-Wuppertal-Houston group used a variation of this method where instead of relying on an expansion, a direct extrapolation of constant entropy density contours from imaginary chemical potential to real values was used, under strangeness neutrality conditions.
This analysis rules out the CP at $\mu_B < 450$ MeV at the $2\sigma$ level~\cite{Borsanyi:2025dyp}.

Holographic approaches provide an alternative handle on the CP. In Ref.~\cite{Hippert:2023bel}, an Einstein-Maxwell-Dilaton model was calibrated to lattice QCD thermodynamics and $\chi_2^B$ at $\mu_B = 0$ using Bayesian inference over two different functional Ans\"atze. Even though the priors allow for the absence of a CP, nearly all posterior samples contain one clustered in the narrow range $T_C \simeq 101$--$108\,\mev$ and $\mu_C \simeq 560$--$625\,\mev$, largely independent of the chosen potential. A complementary V-QCD analysis~\cite{Ecker:2025vnb} employs a Veneziano-limit construction matched at zero temperature with a finite-temperature extension via a van der Waals description of nuclear matter; its Bayesian fit is driven by neutron-star mass--radius and tidal-deformability observations. This astrophysically constrained ensemble yields a strong first-order transition at $T=0$, and favors a CP around $\mu_C \sim 626$ MeV and $T_C \sim 119$ MeV. Despite their different inputs (lattice QCD versus neutron-star observations) and conditions (zero $\mu_S$ and $\mu_Q$ versus $\beta$-equilibrium), both holographic estimates place the CP in the same vicinity of the phase diagram shown in Fig.~\ref{fig:CPplot}.

Functional methods have matured to the point where they provide quantitative estimates for the QCD critical point that are benchmarked to lattice QCD results at $\mu_B = 0$. Predictions within the functional renormalization group (fRG) (Ref.~\cite{Fu:2019hdw}), a hybrid DSE-fRG computation~\cite{Gao:2020fbl}, and a complementary DSE study with explicit pion and sigma backcoupling~\cite{Gunkel:2021oya} all predict a CP at $\mu_C \sim 600$--$650$~MeV and $T_C \sim 110$--$120$ MeV, reflecting the robustness of the functional predictions. Note that the DSE estimate from Ref.~\cite{Gunkel:2021oya} is obtained for $\mu_Q=0$, $\mu_S=\mu_B/3$ instead of $\mu_Q = \mu_S = 0$ conditions in other estimates. 

In summary, various approaches increasingly converge on a rather narrow region for the QCD critical point around $T_C \sim 100$--$120$ MeV and $\mu_C \sim 550$--$650$ MeV. 
At the same time this should not be interpreted as a firm constraint on the CP location, as the various calculations rely on different truncations and input assumptions.
It rather indicates a region where the CP is most likely to be found if it exists. 
Alternative descriptions such as chiral mean-field, model-based constructions tuned simultaneously to heavy-ion data and neutron star observations~\cite{Steinheimer:2025hsr} can accommodate a qualitatively different high-density phase structure, where the CP is located in a cold and dense phase well outside of the region shown in Fig.~\ref{fig:CPplot}. 
The preferred CP region in Fig.~\ref{fig:CPplot} should therefore be viewed as a target band rather than a precise prediction.

\section{Fluctuations and correlations}
\label{sec:fluctuations}
In general, fluctuations and correlations are sensitive to the dynamics and the underlying degrees
of freedom of a system. Textbook examples are energy fluctuations which are characterized by the
heat capacity~\cite{LandauLifshitz:StatPhys}, and most prominently in the context of this review, the long-range correlations close
to a critical point which result in large fluctuations leading, for example, to critical
opalescence~\cite{LandauLifshitz:StatPhys}. Besides exploring the phase diagram, fluctuations and correlations have been used to study
the degrees of freedom in the system. For example, fluctuations of the net charge and baryon number
are sensitive to the fractional charge/baryon number of the quarks in QCD
\cite{Asakawa:2000wh,Jeon:2000wg,Ejiri:2005wq,Parra:2025fse}. 
Also correlations between conserved
charges may be used to test if the system is actually deconfined
\cite{Koch:2005vg,Majumder:2006nq}. For an overview see e.g. \cite{Koch:2008ia,Asakawa:2015ybt}.

As already mentioned in the introduction, fluctuations of conserved charges are sensitive to the
structure of the QCD phase diagram in general and the QCD critical point in particular. The reason
for this is that the cumulants of the conserved charge distribution are related to the derivatives
of the pressure with respect to the chemical potential. 
In particular, for the baryon number
cumulants, $\kappa_{n}[B]$, we have
\begin{equation}
    \kappa_n[B] = \frac{\partial^n (\ln Z)}{\partial(\mu_B/T)^n} = \frac{V}{T} \frac{\partial^n
      P}{\partial(\mu_B/T)^n}\, ,
\end{equation}
One additional feature of cumulants is that they can be directly calculated in thermal field
theories such as lattice QCD. These calculations typically determine so-called susceptibilities, which are defined as
\begin{equation}
  \chi_n[B] = \frac{\partial^n (P/T^4)}{\partial(\mu_B/T)^n}\, ,
  \label{eq:susz_define}
\end{equation}
and are trivially related to the cumulants
\begin{equation}
  \kappa_n[B] = V T^3 \chi_n[B]\,.
  \label{eq:cum_susz}
\end{equation}
At vanishing chemical potential susceptibilities up to $8^{th}$ order have been extracted in lattice QCD \cite{Borsanyi:2018grb,Bazavov:2017dus}. Because of the fermion sign
problem, lattice QCD calculations at finite chemical potential are not possible. However,
susceptibilities can be calculated at imaginary chemical potential, which can then be analytically
continued to real values of the chemical potential \cite{Borsanyi:2018grb}. Alternatively, they may
be obtained via Taylor expansion in powers of the chemical potential. Both methods, however, are
restricted to small values of the baryon number chemical potential, $\frac{\mu_B}{T}\lesssim 3$.

Mathematically, cumulants are best expressed in terms of their generating function 
\begin{align}
g(t)= \ln\left[ \sum_{n}P(n) e^{t\,n}\right] \,,
  \label{eq:cum_gen}
\end{align}
where $P(n)$ is the distribution of the number of a given charge, $n$. The cumulants are then
obtained via
\begin{align}
\kappa_{k}=\frac{\partial^{k}}{\partial t^{k}} \left. g(t)\right|_{t=0} \, .
  \label{eq:cum_from_gen}
\end{align}
Cumulants and their generating function can also be defined for distributions of more than one type
of particle (see e.g. \cite{Holzmann:2024wyd}). For example, the co-variance between two
types of particles $a$ and $b$ is given by
\begin{align}
  {\rm cov}(a,b) = \kappa_{1,1}[a,b]
= \frac{\partial^2}{\partial s \,\partial t}
  \ln \sum_{a,b} P(a,b)\, e^{s a + t b}\bigg|_{s=t=0}\,.
  \label{eq:covariance}
\end{align}

Cumulants are extensive quantities, i.e. they scale with the size/volume of the system. Since in
heavy-ion collisions the size of the system is not well known and controlled, one typically studies
ratios of cumulants to remove the leading volume dependence. This also facilitates the comparison
with the susceptibilities from lattice QCD, as their ratios are the same as those for the corresponding
cumulants. Cumulant ratios typically considered are

\begin{align}
  \frac{\kappa_{2}}{\kappa_{1}}; \;\;\;   \frac{\kappa_{3}}{\kappa_{2}} = S\sigma; \;\;\;
  \frac{\kappa_{4}}{\kappa_{2}} = K\, \sigma^{2} \,
  \label{eq:cum_ratios}
\end{align}
where $S$, $K$, and $\sigma$ are the skewness, kurtosis, and standard deviation, respectively, which
are commonly  used to characterize distributions.

While taking the ratio removes the leading dependence on the volume, there is still the remaining
effect of volume fluctuations. Even for the best centrality cuts the impact parameter of the
collisions and thus the volume of the produced systems changes from event to event~\cite{Das:2017ned}. As we shall discuss
below, these volume fluctuations give rise to significant corrections
\cite{Jeon:1999gr,Jeon:2003gk,Skokov:2012ds} which need to be controlled.

At lower collision energies, where the production of anti-baryons can be neglected, it may be
advantageous to study factorial cumulants $\fc{k}$ instead of cumulants $\kappa_{k}$. Given the
distribution $P(n)$, the factorial cumulant generating function, $g_{F}$, is defined as 
\begin{align}
g_{F}(z)= \ln \left[ \sum_{n}P(n) z^{n} \right] \,
  \label{eq:fac_cum_gen}
\end{align}
and the factorial cumulants are again obtained via differentiation
\begin{align}
\fc{k}=\frac{\partial^{k}}{\partial z^{k}} \left. g_{F}(z)\right|_{z=1} \, .
  \label{eq:fac_cum_from_gen}
\end{align}
The cumulant and factorial cumulant generating functions are related via
\begin{align}
g_{F}(z) = g(\ln(z)) \, ,
  \label{eq:generate_relate}
\end{align}
so that the factorial cumulants can be expressed as a linear combination of regular cumulants and
vice versa
\begin{align}
\fc{k} = \sum_{j=1}^{k}s(k,j)\kappa_{j}; \;\;\; \kappa_{k}=\sum_{j=1}^{k}S(k,j)\fc{j} \, ,
  \label{eq:fac_cum_to_cum}
\end{align}
where $s(k,j)$ and $S(k,j)$ are Stirling numbers of the first and second kind, respectively.
As a result, in the presence of a QCD critical point, factorial cumulants of a given order scale with
the same power of the correlation length as cumulants of the same order.  

One important feature of factorial cumulants is that they represent the integrated genuine
correlation functions, or in other words, they measure the true correlations in the system
\cite{Bzdak:2016sxg}.
Another, related property, is that all factorial cumulants of order $n>1$ vanish for a Poisson distribution,
i.e. factorial cumulants measure the deviation from Poisson statistics.
Cumulants, on the other hand, measure deviation from Gaussian statistics, since $\kappa_{k>2}=0$ for a Gaussian
distribution.
Another useful feature is that factorial cumulants, $\fc{n}\{p\}$, of a distribution which is folded with a
binomial distribution with the Bernoulli probability $p$ are simply related to that of the original distribution, $\fc{n}$, via
\begin{align}
\fc{n}\{p\} = p^{n} \fc{n}.
  \label{eq:fc_binom_scale}
\end{align}
Thus, for small Bernoulli probabilities the factorial cumulants vanish,
$\fc{n}\{p\rightarrow 0\}\rightarrow 0$ for $n>1$, demonstrating that
for small acceptance windows the resulting (factorial) cumulants are consistent with those of a
Poisson distribution.

One disadvantage of factorial cumulants is that they are not directly related to
derivatives of the pressure with respect to the chemical potential, and thus are not easily obtained
in lattice QCD. Also, while possible
to define for net-baryons, factorial cumulants are tedious to work with in this case
\cite{Bzdak:2016sxg}. 

As already pointed out, one important feature of cumulants is that they can be measured in
experiment and calculated in lattice QCD. In principle this allows for a direct comparison of theory and
experiment. However, as we shall elaborate in the next section, some care has to be taken for such a comparison to be meaningful.

\section{Comparing experiment with theory}
\label{sec:exp_vs_theory}
\begin{figure}
\begin{centering}
\includegraphics[width=0.48\textwidth]{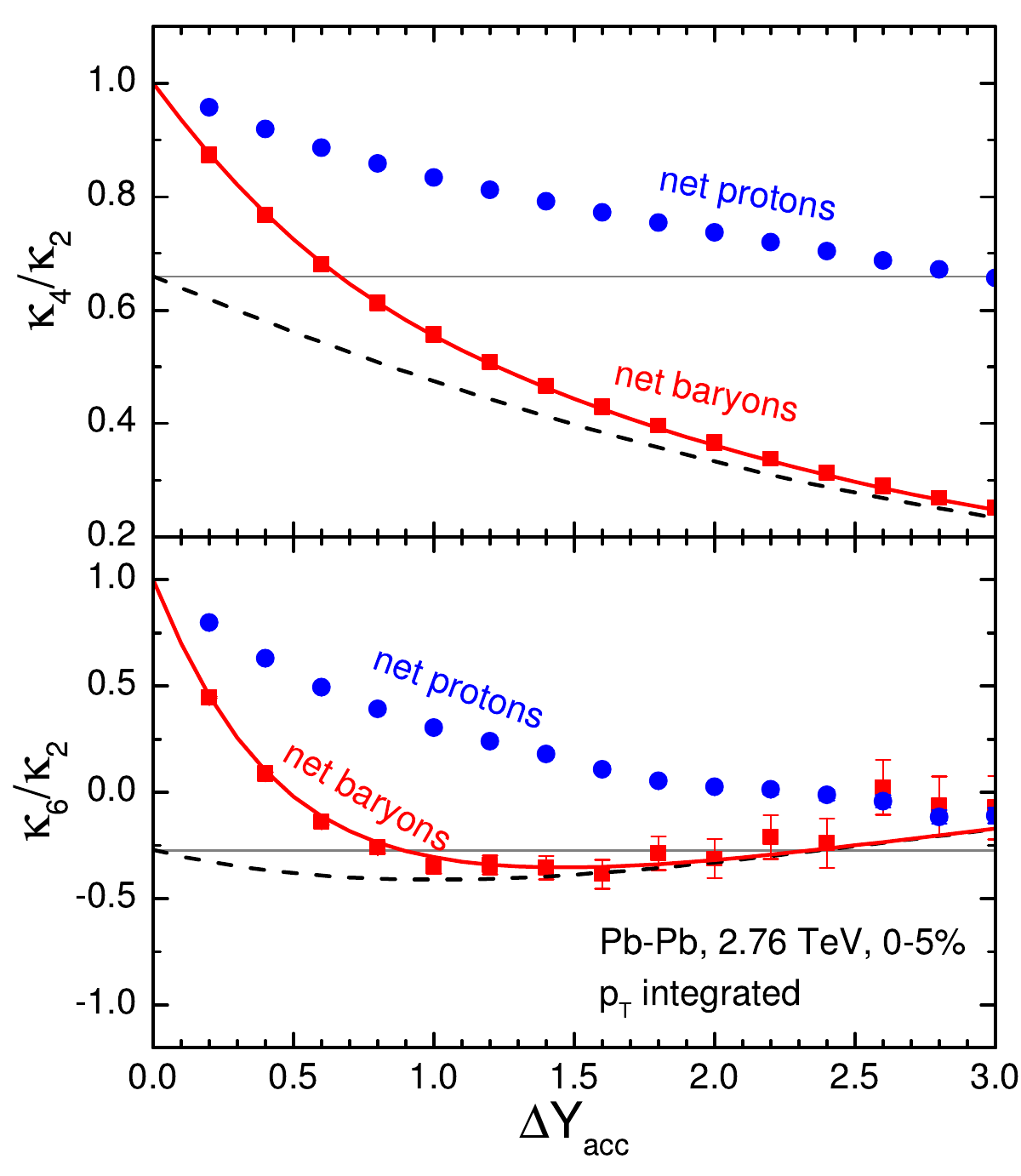}
\includegraphics[width=0.503\textwidth]{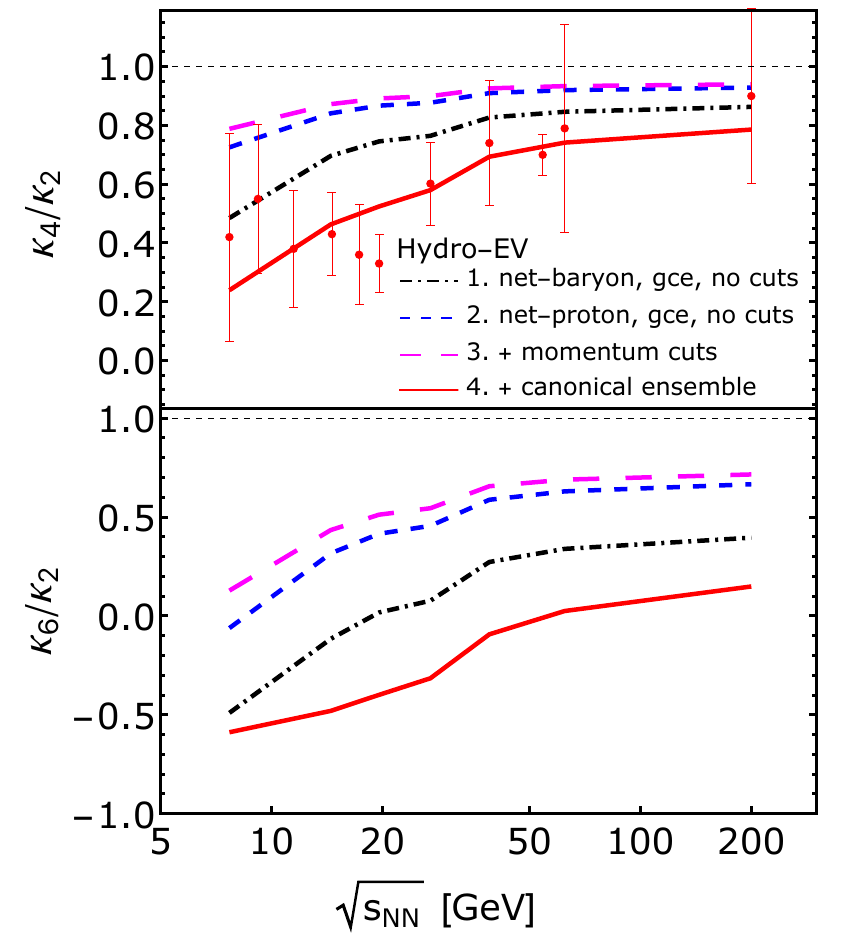}
\par\end{centering}
\caption{Left panel: Cumulant ratio $\kappa_{4}/\kappa_{2}$ (upper panel) and $\kappa_{6}/\kappa_{2}$
(lower panel) as a function of the acceptance window in rapidity,
$\Delta Y$, for a system created in heavy-ion collisions at the LHC.
The horizontal gray lines represent the result from lattice QCD calculations
for the net baryons \cite{Borsanyi:2018grb,Bazavov:2017dus}. The
black dashed lines show the effect of global charge conservation
while the red lines also include thermal smearing. The blue points
are the results for the net-proton cumulant ratio, again with charge
conservation and thermal smearing included. The blue diamonds are
the results for net-proton cumulants using the method of \cite{Kitazawa:2011wh,Kitazawa:2012at}.
For details see \cite{Vovchenko:2020kwg}, where this figure is adapted
from. 
Right panel: Cumulant ratio $\kappa_{4}/\kappa_{2}$ (upper panel) and $\kappa_{6}/\kappa_{2}$ (lower panel) as a function of collision energy for RHIC-BES collider energies for a fixed rapidity window. Hydro-EV model calculations from~\cite{Vovchenko:2021kxx} depicting cumulant ratios of (i) net baryons in the grand canonical ensemble without momentum cuts (dash-dotted black line), 
(ii) the same but for net protons (dashed blue line), (iii) net protons with momentum cuts (dashed magenta line), and (iv) net protons with momentum cuts and baryon number conservation effects included (solid red line).
}
\label{fig:protons_vs_baryons}

\end{figure}

When comparing cumulants measured in experiment with those obtained in thermal field theory
calculations one needs to be aware that the systems probed are different in many important
aspects.
\begin{itemize}
\item \emph{Global charge conservation:} Finite temperature field theory calculations
are commonly done in the grand-canonical ensemble, where the system
can exchange conserved charges with the external (infinite) heat bath.
Thus the charges such as baryon number, $B$, strangeness, $S$, and
electric charge, $Q$, are conserved only on the average. In a heavy
ion collision the charges of the entire system, on the other hand,
are conserved explicitly. While one can mimic a grand-canonical ensemble by considering
only a subsystem, typically by looking only at slices in rapidity
\cite{Koch:2008ia}, effects of global charge conservation remain
since the entire system is still finite. Corrections due to global
charge conservation can be quite sizable \cite{Bzdak:2012an,Braun-Munzinger:2016yjz,Savchuk:2019xfg,Pruneau:2019baa}.
While most estimates of these corrections are based on the (ideal)
hadron resonance gas, meanwhile it has been shown that these corrections
can be calculated for \emph{any} equation of state, in particular
that of QCD \cite{Vovchenko:2020tsr,Vovchenko:2020gne,Vovchenko:2021yen}.
For the commonly used cumulant ratios one finds
\begin{align}
\frac{\kappa_{2}[B]}{\kappa_{1}[B]} & =(1-\alpha)\,\frac{\chi_{2}^{B}}{\chi_{1}^{B}}\label{eq:sam_k21}\\
\frac{\kappa_{3}[B]}{\kappa_{2}[B]} & =(1-2\alpha)\,\frac{\chi_{3}^{B}}{\chi_{2}^{B}}\label{eq:sam_k32}\\
\frac{\kappa_{4}[B]}{\kappa_{2}[B]} & =(1-3\alpha\beta)\,\frac{\chi_{4}^{B}}{\chi_{2}^{B}}-3\alpha\beta\left(\frac{\chi_{3}^{B}}{\chi_{2}^{B}}\right)^{2}.\label{eq:sam_k42}
\end{align}
Here, $\kappa_{n}[B]$, represents the baryon number cumulant of order
$n$, \emph{corrected} for global baryon number conservation. $\chi_{n}^{B}$
denotes the $n^{th}$-order baryon number susceptibility
for a grand-canonical ensemble in full QCD, as for example determined
by lattice QCD. The factor $\alpha$ denotes the fraction of the total
number of baryons plus anti-baryons which is actually observed, $\alpha=\frac{\ave{N_{B}}_{observed}}{\ave{N_{B}}_{4\pi}}$,
and $\beta=1-\alpha.$ Since typically only protons are observed, $\alpha<\frac{1}{2}$.
We note that the expressions Eqs. \eqref{eq:sam_k21}-\eqref{eq:sam_k42}
are valid in the limit where the correlation length is small compared to the
system under consideration. As discussed in detail in \cite{Vovchenko:2020tsr},
this is the case for the systems studied in heavy-ion collisions.  Similar expressions have also been derived for the other conserved charges, $Q$ and $S$, as well as for mixed cumulants \cite{Vovchenko:2020gne}.
The above are minimum corrections due to global charge conservation.
In reality, the corrections can be even larger due to the dynamical nature of heavy-ion collisions.
For this purpose, local charge conservation corrections have also been explored \cite{Braun-Munzinger:2023gsd,Vovchenko:2024pvk}.
\item \emph{Thermal smearing: }The above relations between measured cumulants
and those obtained in the grand-canonical ensemble do not take into
account ``thermal smearing'', i.e. the fact that due to thermal motion
even for a boost invariant system particles in a given spatial rapidity bin are distributed 
over a range in momentum-space rapidities. As a result of the thermal
smearing, the observed cumulants approach the Poisson limit as the
acceptance in rapidity approaches zero~\cite{Ling:2015yau}. 
\item \emph{Baryons vs. protons: }Protons are baryons, but not all baryons
are protons. Thermal field theory calculations can typically only
calculate baryon number susceptibilities as they are associated with
the derivative of the pressure w.r.t the baryon number chemical potential.
Experiments, on the other hand, usually cannot  measure neutrons and
are thus restricted to net-proton number cumulants. As argued in Refs.
\cite{Kitazawa:2011wh,Kitazawa:2012at}, in the presence of many pions,
charge exchange reactions effectively randomize the proton and neutron
numbers. In this case, the proton cumulants can be obtained from the
baryon-number cumulants by a binomial folding with a Bernoulli probability
of $p=\ave{N_{p}}/\ave{N_{B}}\simeq1/2$.
As discussed in the previous
section such a binomial folding moves the cumulants closer to the
Poisson (Skellam) limit.
\item \emph{Volume fluctuations:} As already alluded to in the previous section, the event-by-event fluctuations of
  the impact parameter and thus system size cannot be totally removed even with the best centrality
  selection. Since (factorial) cumulants scale with the system size (volume) their values also fluctuate. While
  the dependence on the mean system size can be removed by taking ratios of cumulants, 
  volume fluctuations still affect the measured cumulants \cite{Skokov:2012ds} and thus need to be
  understood and, if possible, be removed. The STAR collaboration applies so-called centrality bin
  width corrections (CBWC) \cite{Luo:2013bmi} in order to suppress volume fluctuations. As discussed in
  \cite{Friman:2025ulh}, the CBWC procedure indeed is able to reduce the effect of volume fluctuations
  but in some cases may even over-correct the results, thus affecting the physics. Unfortunately,
  so far no criterion has been established quantifying the quality of the correction. Another recently proposed method
  utilizes mixed events to extract the contribution from volume fluctuations
  \cite{Rustamov:2022sqm}. Also in this case not all effects may be removed. However, as discussed
  in \cite{Holzmann:2024wyd}, this method provides an estimate for the bias, which can be
  constrained by the systematics of the system under investigation. One can also show that this bias is
  parametrically suppressed for systems with large charged-particle multiplicities, such as those
  generated in LHC-energy collisions. Unfortunately, this is not the case for the energies where a
  possible QCD critical point is expected to be found. Therefore, one needs to either rely on
  simulations or develop and measure so-called strongly intensive observables
  \cite{Gorenstein:2011vq,Sangaline:2015bma} which are not affected by volume fluctuations. Also, a new method based on the Edgeworth expansion \cite{Wang:2025fve} has recently been proposed which is claimed to be able to determine the cumulants without a specific centrality selection.   

\end{itemize}
The effect of the first three corrections, baryon number conservation, thermal smearing and protons
vs baryons, is illustrated in Fig. \ref{fig:protons_vs_baryons}
where in the left panel we show the dependence of the cumulant ratios $\kappa_{4}/\kappa_{2}$
and $\kappa_{6}/\kappa_{2}$ as a function of the size of the rapidity
acceptance window for a typical system produced at LHC energies (for
details, see \cite{Vovchenko:2020kwg}). Here the horizontal gray
lines represent the value for the cumulant ratio as obtained from
lattice QCD \cite{Borsanyi:2018grb,Bazavov:2017dus}. The black dashed
lines show how this cumulant ratio changes with $\Delta Y$ due to
global charge conservation. The red lines are the result for the cumulant
ratio if both charge conservation and thermal smearing are taken into
account. One sees that, due to thermal smearing, the cumulant
ratio approaches the Poisson limit of $\kappa_{4}/\kappa_{2}=1$ as
the acceptance window becomes small. Finally, the blue points show
the cumulant ratio for net-protons instead of net baryons with both
charge conservation and thermal smearing included. The blue diamonds
are the net proton cumulants obtained using the method of \cite{Kitazawa:2011wh,Kitazawa:2012at}.
The blue points are what an experiment such as ALICE is expected to observe if
the system created is in thermal equilibrium and if there are no effects
other than the fluctuations predicted by lattice QCD. For both cumulant
ratios, we see a substantial difference between the predicted value
from lattice QCD and what is measured in the experiment using net protons.
In particular, for the hyper-kurtosis, $\kappa_{6}/\kappa_{2}$, lattice QCD
predicts a negative value while that for net protons turns out to be positive. A negative
sign of the hyper-kurtosis has been argued to be a signal for the remnant
of chiral criticality \cite{Friman:2011zz}. Therefore, great care
needs to be taken to reveal the underlying baryon cumulants from those
measured. Such an endeavor will likely require the measurement of several cumulant
ratios as a function of the size of the acceptance window in order to minimize the systematic uncertainties. 
Only second-order proton number cumulants have been fully measured so far~\cite{ALICE:2019nbs,ALICE:2022xpf}.

The right panel of Fig. \ref{fig:protons_vs_baryons} shows the same cumulant ratios as a function of collision energy for Au-Au collisions at RHIC as evaluated within the hydrodynamic model calculations from~\cite{Vovchenko:2021kxx}. 
The figure first shows net-baryon cumulant ratios in the grand-canonical ensemble without momentum cuts (dash-dotted black line), which exhibit a suppression relative to the Skellam baseline value of unity and reflect correlations due to the baryon excluded volume, in line with lattice QCD susceptibilities.
This is the type of prediction one can obtain from thermal field theory calculations.
However, if one considers net protons instead of net baryons (dashed blue line), the cumulant ratios move significantly closer to the Skellam baseline, reflecting the dilution of baryon correlations due to missing neutrons. 
Correlations are further diluted by momentum cuts (dashed magenta line).
Finally, when canonical effects are included (solid red line), the cumulant ratios of net protons move significantly away from the Skellam baseline, with the final result being closer to those of net baryons in the grand-canonical ensemble without momentum cuts.
This interplay of different effects may explain the fair agreement between net-baryon susceptibilities from lattice QCD (grand-canonical, no momentum cuts) and the measured net-proton cumulant ratios (canonical, momentum cuts) even though they correspond to different observables (baryons vs protons).
These results suggest that comparisons between lattice QCD susceptibilities and experimental cumulant ratios should be made with caution, and that directly equating the two may lead to misleading conclusions.

In addition to the aforementioned issues one should also be aware that
the systems created in heavy-ion collisions are dynamic, i.e. they
evolve with time whereas the systems studied in thermal field theories
are static and in thermal equilibrium. Of course, if the time evolution
of the system created in heavy-ion collisions is governed by hydrodynamics
and the typical hydrodynamic scale is larger than the correlation
length responsible for critical fluctuations, as argued e.g. in \cite{Stephanov:2017ghc},
then the application of (local) thermal equilibrium may be a reasonable
approach. If one wants to calculate the effect of critical fluctuations,
diffusion and non-hydro modes need to be propagated as well. This
can be done either via stochastic hydrodynamics \cite{Landau_Hydro} or
by explicitly propagating two- and higher-order critical correlation
functions as proposed in \cite{Stephanov:2017ghc}. 

\begin{figure}
\begin{centering}
\includegraphics[width=0.5\textwidth]{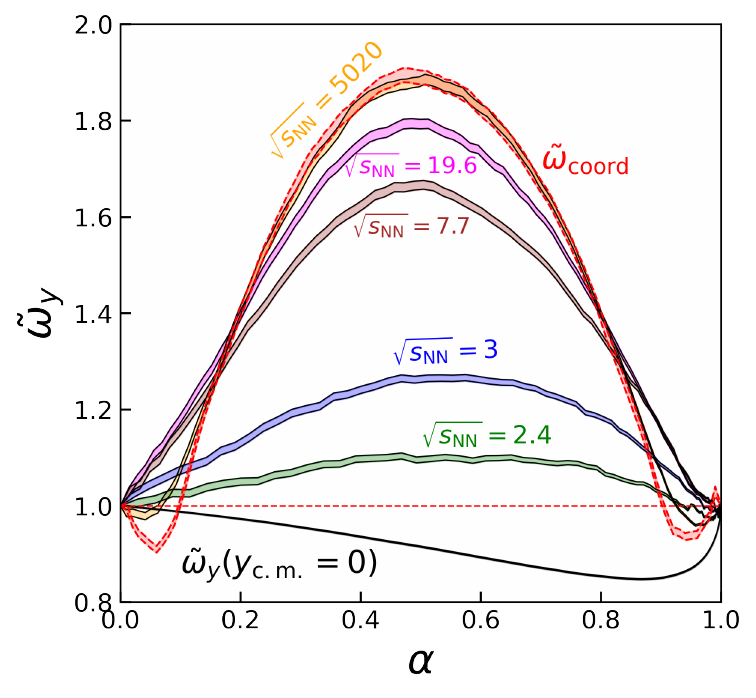}
\par\end{centering}
\caption{Corrected scaled variance $\tilde \omega_{y}$ of particle number in rapidity acceptance as
  a function of the fixed acceptance fraction $\alpha_y$, which is the ratio of accepted to total
  number of particles.  Calculations are performed for a system  of $N = 400$ particles at $T = 1.06
  T_c$ and $n = 0.95 n_c$. Different bands correspond to different magnitudes of the collective flow
  corresponding to the collision energies in a Bjorken picture.  The limiting cases of coordinate,
  red band, labeled $\tilde \omega_{\rm coord}$, and rapidity acceptance, black line, labeled
  $\tilde \omega_y (y_{\rm cm} = 0)$, in the absence of collective expansion are also shown. For details, see \cite{Kuznietsov:2024xyn} where this figure is adapted from. }
\label{fig:md_vs_sqrts}
\end{figure}

At lower collision energies, which correspond to systems at higher net
baryon density but lower energy density, non-equilibrium effects are
expected to become relevant, so that approaches based on hydrodynamics
may no longer be reliable. Instead one has to resort to some kind of
kinetic theory, which has not yet been developed for QCD matter. However,
in order to develop some intuition about the importance of non-equilibrium
effects and the possibility to detect signals for a dynamical system,
it may be a good first step to study classical molecular dynamics.
This has been recently done in Refs. \cite{Kuznietsov:2022pcn,Kuznietsov:2024xyn}
for a Lennard-Jones fluid which does have a critical point in the same universality class as the conjectured QCD critical point. This study also addressed, at least qualitatively,
another important difference between theory and experiment:

\begin{itemize}
\item \emph{Theory calculates in coordinate space while experiment measures in
momentum space: }In thermal field theory one works in the grand-canonical
ensemble. In practice this means that one considers a system with \emph{spatial
}sub-volume $V_{S}$ of a large total volume $V_{T}$ such that $V_{S}\ll V_{T}$.
The thermodynamic limit then corresponds to the limit where both volumes
go to infinity, $V_{S},\,V_{T}\rightarrow\infty$ while still preserving
that $V_{S}\ll V_{T}$. Let us, therefore, consider the situation where
$V_{T}$ is large but not infinite. In the limit of $V_{S}\ll V_{T}$
but $V_{S}\gg\xi^{3}$ one recovers, after suitable corrections for
global charge conservation as discussed above, the grand-canonical
results for the cumulants. Here $\xi$ denotes the relevant correlation
length. Thus, in theory one studies the fluctuations of a small \emph{spatial
}sub-volume which does particle and energy exchange with the large
total volume. At the same time one integrates over all particle momenta
in the small sub-volume. In experiment the situation is just the
opposite: One studies the cumulants of a small sub-volume in \emph{momentum
}space characterized by, for example, cuts in rapidity. At the
same time, experimental measurements integrate over all coordinate
space. This can lead to quite different results as demonstrated in
\cite{Kuznietsov:2022pcn}. To see this, let us consider a non-relativistic
system, such as the Lennard-Jones fluid which is governed by a two-particle
interaction in coordinate space, $V(x_i,x_j) = V(|x_{i}-x_{j}|)$. The Hamiltonian
of such a system is 
\begin{equation}
H=\sum_{i}\frac{p_{i}^{2}}{2m}+\sum_{i,j}V(x_i,x_j)
\end{equation}
so that the partition function for a system of $N$ particles in a total
phase-space volume $\Omega=\Delta P \times \Delta R$ is given by
\begin{align}
Z & =\int_{\Omega}dx_{1}dp_{1}\cdots dx_{N}dp_{N}\exp(-\frac{H}{T})\nonumber \\
 & =\int_{\Delta P}dp_{1}\cdots dp_{N}\exp\left(-\frac{\sum_{i}p_{i}^2}{2mT}\right)\nonumber \\
 & \times \int_{\Delta R}dx_{1}\cdots dx_{N}\exp\left(\frac{-\sum_{i,j}V(x_{i},x_{j})}{T}\right)\nonumber \\
 & =Z_{P}Z_{R}
\end{align}
Obviously, the partition function factorizes in a spatial, $Z_{R}$,
and momentum, $Z_{P}$, piece with 
\begin{align}
Z_{R} & =\int_{\Delta R}dx_{1}\cdots dx_{N}\exp\left(-\frac{\sum_{i,j}V(x_{i},x_{j})}{T}\right)\\
Z_{P} & =\int_{\Delta P}dp_{1}\cdots dp_{N}\exp\left(-\frac{\sum_{i}p_{i}^2}{2mT}\right).
\end{align}
If we integrate over all momenta but limit the size of the spatial
volume, as it is done in theory, we study the behavior of $Z_{R}$ and 
are sensitive to the correlations introduced by the interaction.
If, on the other hand, we limit the momentum space but integrate over
the entire spatial volume as it is done in experiment, the resulting
partition function $Z\sim Z_{P}$ is essentially that of a gas of
non-interacting particles. Therefore, one will not observe any non-trivial
correlations and fluctuations. Exactly this has been demonstrated in
Refs.~\cite{Kuznietsov:2022pcn,Kuznietsov:2024xyn} by explicit molecular dynamics calculations of the Lennard-Jones liquid. Thankfully, the systems created in heavy-ion collisions
are not static but exhibit considerable collective flow, especially
at high energies. Therefore, momentum space and coordinate space
are correlated, and cuts in momentum space correspond to some cuts 
in coordinate space. However, at lower energies, where the collective
flow is rather modest, one should expect that the signals will become
weaker simply because one is approaching the above discussed static
limit. This behavior is seen in explicit calculations of \cite{Kuznietsov:2024xyn}, where a Bjorken-
like longitudinal flow profile has been
superimposed on the molecular dynamics simulations. In Fig.~\ref{fig:md_vs_sqrts} we show the scaled
variance as a function of the momentum-space acceptance for different strengths of the collective
flow profile labeled by the corresponding collision energy in a simple Bjorken picture. Here,
the scaled variance $\tilde{\omega}$ is corrected for global particle number conservation using the
procedure discussed above [cf. Eq.~\eqref{eq:sam_k21}]. The signal without flow (black line)
corresponds to that of a non-interacting gas in the micro-canonical ensemble
\cite{Kuznietsov:2022pcn}. The signal obtained in coordinate space, which would correspond to predictions from typical theory calculations, is shown as the red line. We see that with increasing flow or increasing collision energy the scaled variance increases and reaches that obtained from coordinate space cuts. However, at the lower energies, $\sNN \simeq 3$--$7 \gev$, where the QCD critical point is predicted to be located, the signal is considerably reduced.

\end{itemize}

\section{Non-critical baseline and experimental data }
\label{sec:data_and_predictions}
Ideally one would have a theoretical model which can describe the entire dynamical evolution
of the systems created in heavy-ion collisions, including effects of phase transitions. However, at present such a model is not available, although
developments towards this end are under way
\cite{An:2021wof,Stephanov:2017ghc,Pradeep:2022mkf}. But even with the availability of such a model,
it is good practice to develop a null hypothesis. In other words, one needs a baseline 
which contains all the known physics but does not include correlations associated with a QCD critical
point or a phase transition. Deviations from such a non-critical baseline will then reveal at what energy possible new physics may
be found. Such a baseline should of course include all the corrections
discussed in the previous section such as baryon number conservation etc. Ideally, it should also
reproduce all other observables not associated with a phase transition, such as particle spectra etc. There are several versions
of such a baseline in the literature, none of which unfortunately takes all non-critical effects into account.   
The STAR collaboration \cite{STAR:2021iop,STAR:2022etb,STAR:2025zdq} typically uses
the UrQMD event generator for this purpose. UrQMD conserves all the charges, such as baryon number,
and, being based on kinetic theory, includes the effects of thermal
smearing. In addition, it provides results for (net) proton cumulants in addition
to (net) baryon cumulants. Also, with UrQMD being an event generator one can
apply the same acceptance cuts and centrality selection criteria as
in the experiment. The latter may help to simulate the effect
of volume fluctuations \cite{STAR:2022etb}. We note, that STAR also applies the same centrality bin
width corrections to the UrQMD results as it does to the data. Another approach \cite{Braun-Munzinger:2020jbk} uses
the hadron resonance gas model including global charge conservation
effects and experimental data to constrain the fraction of baryons
in the acceptance. Since this approach is based on an ideal gas of
hadrons, thermal smearing is automatically included, and the model
also provides results for (net) protons. A third approach \cite{Vovchenko:2021kxx} uses
viscous hydrodynamics tuned to reproduce the experimental data for spectra
etc.\ for the time evolution. The particlization is carried out with a 
method which respects global baryon number conservation also for (net)
protons \cite{Vovchenko:2021yen} and, by construction, includes
the effects of thermal smearing. In addition, sampling is done such that
the resulting cumulants agree with those from lattice QCD at vanishing
chemical potential. This is achieved by introducing an excluded volume
correction into the hadron resonance gas equation of state tuned
to reproduce the lattice cumulants. Using an excluded volume is justified
by an analysis of lattice results for the fugacity expansion of the
pressure \cite{Vovchenko:2017gkg,Vovchenko:2017xad}. However, both 
this approach and that based on the hadron resonance gas presently do not account for volume
fluctuations. This may not be such a problem since the STAR data contain centrality bin width
corrections which, as discussed above, remove some effects of volume fluctuations (at least at higher energies), albeit not in a
controlled fashion \cite{Friman:2025ulh}.

\begin{figure}[h]
\centering
\includegraphics[width=0.99\textwidth]{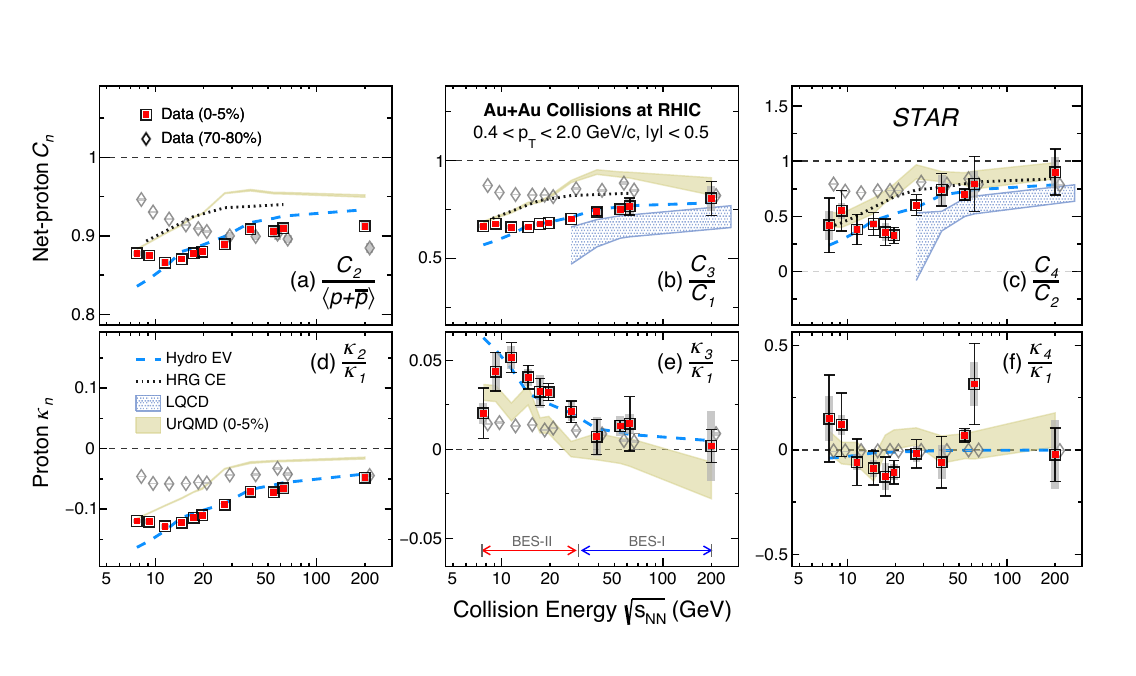}
\caption{Cumulants (top row) and factorial cumulants (bottom row) obtained by the STAR collaboration
  from the second phase of the RHIC beam energy scan \cite{STAR:2025zdq}. Note that, contrary to
  common practice, STAR uses $C_{n}$ to denote cumulants and $\kappa_{n}$ to denote factorial cumulants. Also shown are the
  baselines of \cite{Vovchenko:2021kxx} (blue dashed line, denoted as Hydro-EV in the text),  \cite{Braun-Munzinger:2020jbk} (dotted
  black line, denoted as HRG-CE in the text) as well as UrQMD calculations by STAR (brown band).  Also shown are lattice QCD
  results for net baryons uncorrected for global baryon number conservation
  \cite{Bazavov:2020bjn}. Figure adapted from \cite{STAR:2025zdq}}\
\label{fig:star_BES2}
\end{figure}

In Fig.~\ref{fig:star_BES2} we show the comparison of the recent STAR data from RHIC BES-II for
(factorial) cumulants with those baselines. We note that both the result based on hadronic resonance
gas (HRG-CE)
\cite{Braun-Munzinger:2020jbk} (dotted black line) and that based on hydrodynamics with eigenvolume corrections
\cite{Vovchenko:2021kxx} (blue dashed line) were obtained prior to the data. They apply only to the central data (red
squares). All three baselines describe the trend of the data as a function of the collision energy
rather well, and the result from hydrodynamics with eigenvolume (Hydro-EV) corrections even agrees (within
errors) quantitatively with the measurement {\em except} for the lowest two collision
energies. For the lowest energies, however, the data show some non-monotonic change while the Hydro-EV baseline continues to
decrease (increase) for the second (third) order factorial cumulants. This trend in the data
actually seems to continue to lower energies, where the STAR collaboration finds even larger (smaller)
second (third) order factorial cumulants at a collision energy of $\sNN =3 \gev$ \cite{STAR:2022etb}.

We note that the essential difference between the Hydro-EV and the HRG-CE baseline is the
eigenvolume correction in the former. This correction actually corresponds to a short-range repulsion
among the baryons. Thus one may speculate \cite{Vovchenko:2021kxx} that at lower energies this repulsion needs to disappear or, more
interestingly, it needs to be compensated by an additional attraction in order to agree with the
lowest two energy points, where the HRG-CE agrees much better with the cumulants. This idea has
recently been implemented in a model calculation \cite{Friman:2025swg}, demonstrating that with an
appropriate choice of interaction one could reproduce the observed energy dependence of the (factorial)
cumulants. The need for an attractive interaction could be a first hint of critical dynamics~\cite{Reif:StatPhys}.
However, it could also simply be an effect of the well-known nuclear interactions, which are known to be
a combination of short-range repulsion and long-range attraction, and are responsible for the well-known
liquid-gas phase transition and critical point.
In fact, model calculations of equilibrium fluctuations along the freeze-out line indicate that the contribution of the nuclear liquid-gas phase transition to the cumulants may be sizable at low and intermediate energies~\cite{Mukherjee:2016nhb,Vovchenko:2017ayq,Sorensen:2020ygf}.

\begin{figure}[h]
\centering
\includegraphics[width=0.7\textwidth]{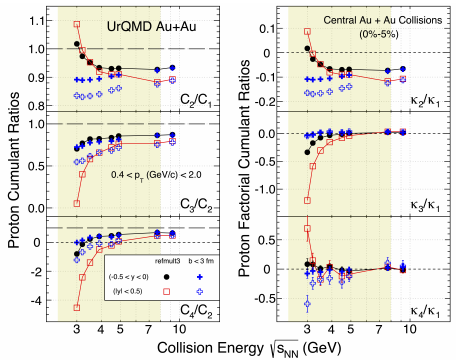}
\caption{Energy dependence of cumulants (left panel) and factorial cumulants (right panel) obtained from UrQMD
  simulations \cite{Zhang:2025ale} for different rapidity acceptance windows and for fixed impact
  parameter (blue crosses) and centrality selection a la STAR (red square and black filled circle). 
  Figure adapted from \cite{Zhang:2025ale}.}
\label{fig:urqmd_volume}
\end{figure}

However, there may be another, more mundane effect which does not involve any extra dynamics. In a recent study \cite{Zhang:2025ale}
the authors used the UrQMD model {\em without} (mean-field) interactions to calculate the energy dependence of (factorial) cumulants for two
cases: (i) With limited impact parameter range ($b<3 \,\rm fm$) and (ii) using the centrality selection
applied in the STAR analysis. The results are shown in Fig.~\ref{fig:urqmd_volume}.  Concentrating
on the second and third order factorial cumulants for symmetric acceptance ($-0.5<y<0.5$) shown in
the right panel, we see that for fixed impact parameter (filled blue crosses) there is a rather mild to
non-existing energy dependence. However, when using the same centrality selection as done for the
experimental data (red open squares) we find that the second  order factorial cumulant rises towards
lower energies and the third order drops, similar to the trend in the  experimental data.  In both cases the authors applied centrality bin width corrections  (CBWC), just as it is done
for the data.
As shown in Ref.~\cite{Zhang:2025ale}, for collisions with limited impact parameter range $(b < 3 \, \rm fm) $ the "volume" or rather participant fluctuations 
are considerably smaller and they show a similar mild energy dependence as the two baselines discussed above (HRG-CE and
Hydro-EV) which do not account for volume fluctuations. Therefore, it may very well be that the rather significant energy dependence seen in
the experimental data may actually be (largely) due to volume fluctuations, which are clearly present in the
STAR centrality selection and which are not fully removed by the CBWC procedure. For higher energies $\sNN \gtrsim 7 \gev$, where the charged particle multiplicity is large, the difference between centrality selections disappears. This is consistent with the finding of \cite{Friman:2025ulh}, which shows that the CBWC procedure seems to successfully remove the effect of volume fluctuations for large multiplicities. Thus, the deviation from the baselines at $7.7\gev$ and $9 \gev$ may indeed indicate the onset of attractive interaction. However, this needs
further investigation before any conclusions about additional, potentially critical dynamics can be drawn, in particular in view of the upcoming CBM experiment which will measure in the region where the QCD critical point is predicted to be located.

\section{Open issues and next steps}
\label{sec:open_issues}
\begin{figure}[h]
\centering
\includegraphics[width=\textwidth]{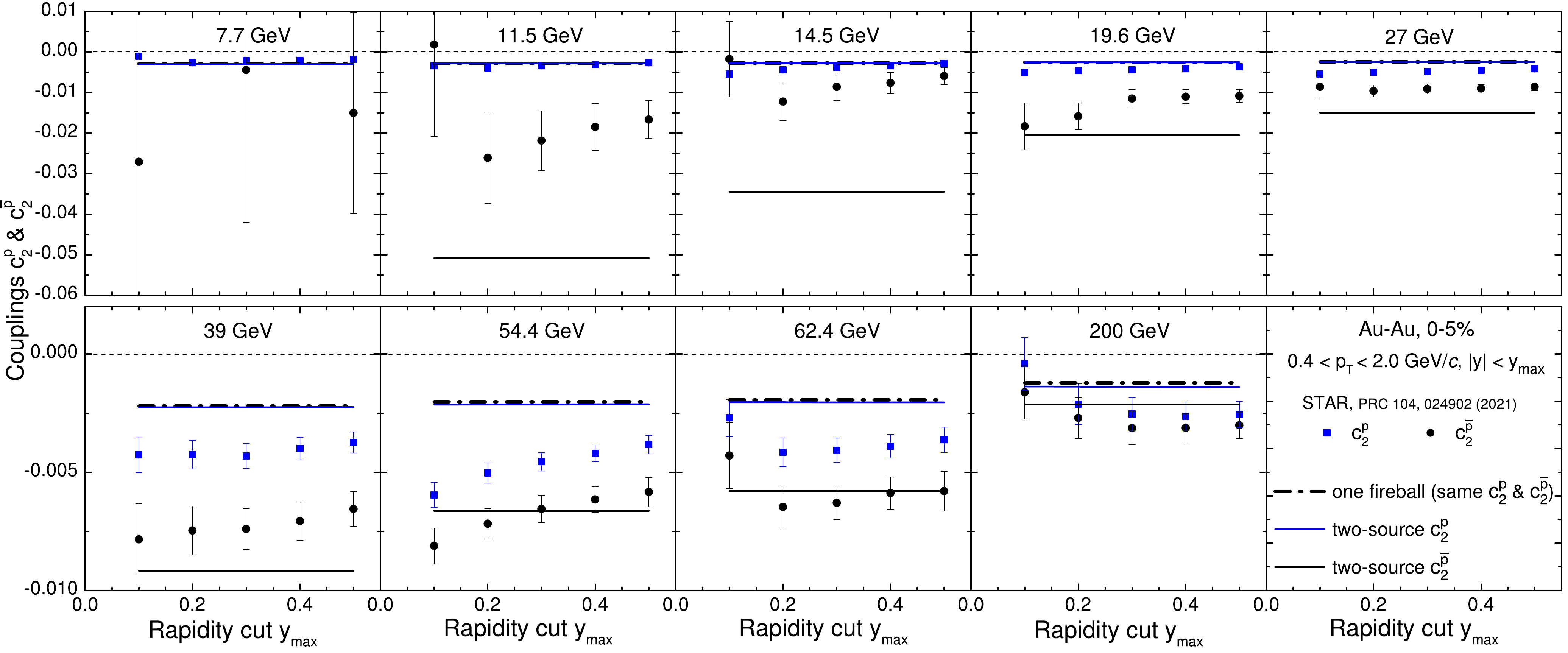}
\caption{Rapidity dependence of the reduced correlation coefficient \cite{Bzdak:2025rhp} for various
  collision energies. The black dashed-dotted line represents the result obtained with a single fireball for both protons and antiprotons. The results for the two-source model are shown as thin blue and black lines for protons and antiprotons, respectively.   
  Figure adapted from \cite{Bzdak:2025rhp} }
\label{fig:bzdak_ratio}
\end{figure}

\begin{figure}[h]
\centering
\includegraphics[width=0.5\textwidth]{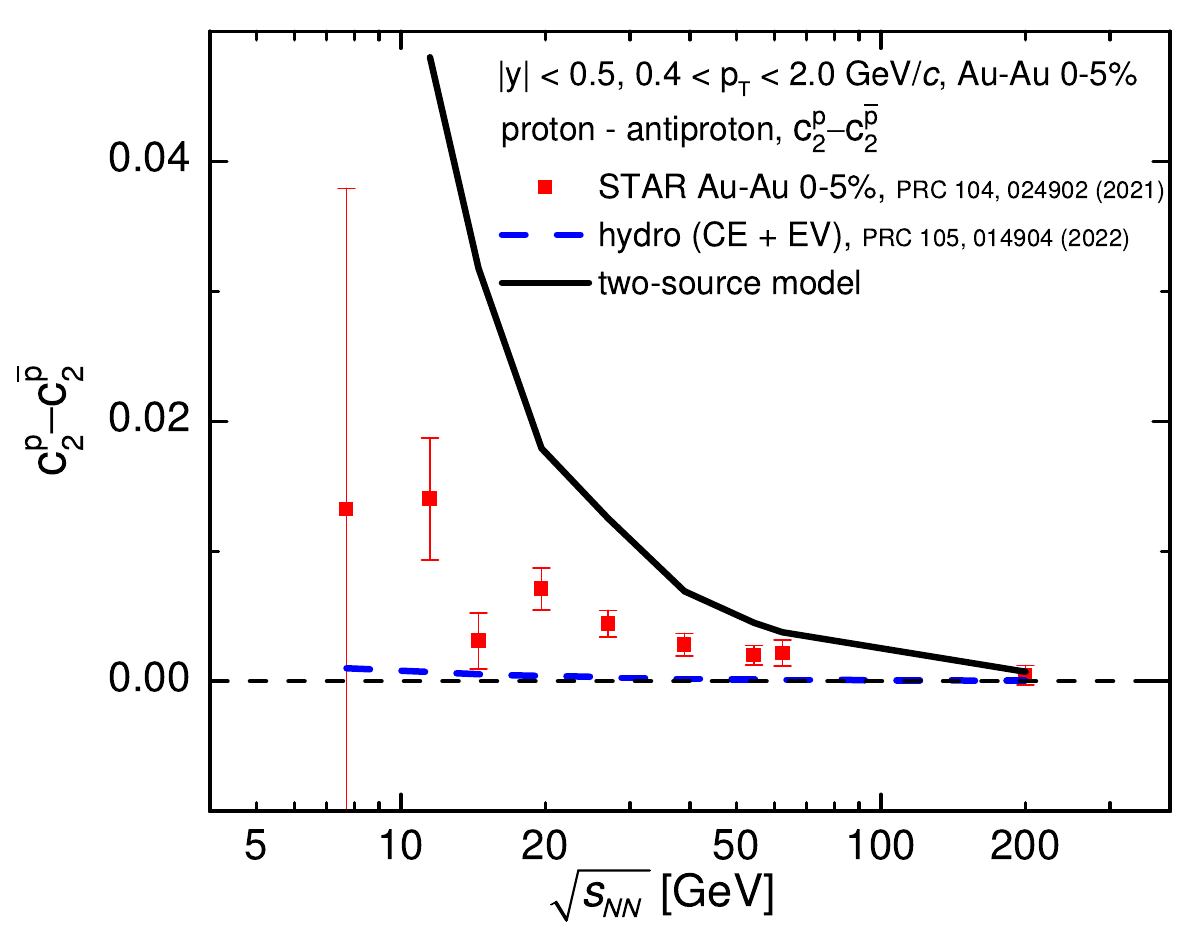}
\caption{ Energy dependence of the difference of  reduced correlation coefficients for protons and
  antiprotons \cite{Bzdak:2025rhp} for the Hydro-EV baseline \cite{Vovchenko:2021kxx} (blue dashed line) and from the two-source model (black line). Figure adapted from \cite{Bzdak:2025rhp} }
\label{fig:bzdak_p_pbar}
\end{figure}

The various baselines discussed in the previous section give the correct trend and in one case even
a quantitative agreement with the energy dependence of the measured (factorial) cumulants for
energies above $\sNN \gtrsim 10\gev$. These
baselines take into account the essential ``trivial'' effects discussed in Sec.~\ref{sec:exp_vs_theory}, baryon number
conservation and the fact that only protons, and not all baryons, are measured and, in the case of UrQMD,
also volume fluctuations. In order to see if there is potentially additional physics it would be
good to provide an additional observable which tests these baselines and their
assumptions. Such an observable has been recently put forward in \cite{Bzdak:2025rhp}. Specifically,
the authors propose to look at the acceptance dependence on the reduced correlation coefficients or couplings
defined as \cite{Bzdak:2017nyp,Bzdak:2016jxo,Bzdak:2017ltv}
\begin{align}
  \cpl{n} = \frac{\fc{n}}{\ave{N}^{n}} 
  \label{eq:coupling_def}
\end{align}
where $\ave{N}=\kappa_{1}=\fc{1}$ is the mean number of protons. Since  global baryon number
conservation, protons vs baryons, as well as volume fluctuations lead only to global and thus long
range correlations, Ref.~\cite{Bzdak:2025rhp} shows that in that case the  reduced correlation coefficients will
be constant as a function of
the size of the rapidity window. This can be seen as follows. For simplicity, let us only consider
correlations in rapidity, $y$. Recall, that the factorial cumulants $\fc{n}$ represent
integrals over the genuine correlations functions, $c_{n}(y_{1},\cdots,y_{n})$ \cite{Bzdak:2016sxg}. Thus we  have for the
second and third order factorial cumulants
\begin{align}
  \fc{2} = \int_{\Delta Y}\int_{\Delta Y}dy_{1}dy_{2}\,c_{2}(y_{1},y_{2}); \;\;\; \fc{3} =
  \int_{\Delta Y}\int_{\Delta Y }\int_{\Delta Y }dy_{1} dy_{2} dy_{3}\,c_{3}(y_{1},y_{2},y_{3}); 
\end{align}
where $\Delta Y$ denotes the acceptance window in rapidity. Since we have only long range correlations, the
genuine correlation functions are constant over the acceptance window,  $c_{2}(y_{1},y_{2})=const.$,
$c_{3}(y_{1},y_{2},y_{3})=const.$ Hence the factorial cumulants scale with the size of the
acceptance window
\begin{align}
\fc{2} \sim (\Delta Y)^{2} \sim \ave{N}^{2}; \;\;\; \fc{3} \sim (\Delta Y)^{3} \sim \ave{N}^{3} 
  \label{eq:fc_scaling}
\end{align}
so that the couplings, $\cpl{n}$ are constant as a function of the acceptance.
In addition to the above scaling of the factorial cumulants, the ratio of the  second order reduced
correlation coefficients for protons and anti-protons are found to be virtually identical for an
ideal hadron gas in the grand canonical ensemble, which underlies both the HRG-CE and Hydro-EV baselines,
$\cpl{2}[p]=\cpl{2}[\bar{p}]$. In Fig.~\ref{fig:bzdak_ratio} we show the results for the rapidity
dependence of the couplings as measured by the STAR Collaboration during the first
phase of the RHIC beam energy scan, BES I, for protons (blue symbols) and anti-protons (black
symbols). We also show, as black dashed dotted line, the results obtained from a single fireball model for protons and antiprotons which are identical. This model agrees very well with the  baseline of  \cite{Vovchenko:2021kxx} without eigen-volume correction. However, the eigen-volume
corrections lead only to small differences, see Fig.~3 of Ref.~\cite{Bzdak:2025rhp}. Within the rather
large errors, the data are consistent with the expected scaling, i.e. they are constant as a
function of acceptance (with the possible exception of $\sNN=50\gev$). Also the reduced proton correlation coefficient is reproduced quantitatively  for energies up to
$\sim 27\gev$. However, contrary to
expectations, the data clearly show a significant difference between the reduced 
correlation coefficients of protons and anti-protons, except for the highest energy $\sNN = 200
\gev$.   Should the new, high statistics data from BES II confirm these results, there is
clearly a need to either  revise the baseline(s) or understand possible new physics which is beyond 
baryon number conservation etc. One such idea put forward in  Ref.~\cite{Bzdak:2025rhp} was a simple
two-source model, which differentiates between produced protons and anti-protons and protons that are
stopped from the target and projectile nuclei. This model leads to the observed increase in the
difference between the coupling for protons and anti-protons,  $\cpl{2}[p]-\cpl{2}[\bar{p}]$, but
quantitatively overpredicts the data (see Fig.~\ref{fig:bzdak_p_pbar}). Certainly such a model is too simple, but it may suggest that
the picture of one fireball may be too simplistic for collisions at lower energies where the
contribution from stopped protons is significant.

The above discussion illustrates the need for more differential data from BES II. For example, the second order (factorial) cumulants are related to integrals of proton-proton rapidity correlations \cite{Bialas:1999tv}, such as the ones measured by the STAR collaboration \cite{STAR:2019cqg} during BES I. Just as for the reduced correlation coefficients one would expect that effects due to volume fluctuations will not affect the shape of these correlation functions. Of course, one needs to assure that the experimental conditions such as acceptance cuts, efficiency corrections etc. are identical for both measurements for such a comparison to be meaningful. The same information can also be extracted from balance functions \cite{Bass:2000az,Jeon:2001ue,Pruneau:2022mui} with similar caveats about experimental conditions. Furthermore it would be desirable to have data for all energies available from the fixed target mode, even though their interpretation will be more difficult since they will not have symmetric rapidity acceptance compared to those from collider measurements. This is not only important to understand the present results and their systematics but it will also be crucial in preparation for the results from the upcoming CBM experiment. 

At lower beam energies an additional complication arises because there is no longer a well-defined boost-invariant mid-rapidity region. 
The net-baryon rapidity distribution is steep and the local thermodynamic parameters vary significantly across the available acceptance even under a local equilibrium scenario. 
Furthermore, spectators and fragmentation remnants can contaminate the measured cumulants. 
This makes it more difficult to interpret measurements of cumulants in terms of equilibrium expectations, calling for more differential analyses in rapidity together with a careful treatment of spectators in both data and models.

Of course there are other observables which should be studied such as transverse momentum fluctuations~\cite{Heiselberg:2000fk,STAR:2005vxr},
light-nuclei yields and their fluctuations~\cite{Sun:2017xrx,DeMartini:2020anq}, electromagnetic observables~\cite{Akamatsu:2025axh}, and finite-size scaling analyses~\cite{Lacey:2014wqa,Sorensen:2024mry}. 
A coherent picture emerging simultaneously from several such observables would significantly strengthen any claim of critical behavior.
Also, if indeed the QCD critical point is located as predicted one would expect that for collision energies below $\sNN \simeq 5 \gev$ the system should cross the first order transition, and thus spinodal breakup should occur. Even though Refs. \cite{Steinheimer:2013xxa,Steinheimer:2019iso} were not able to identify a significant signal, further, more innovative approaches such as advanced machine learning techniques may very well be more successful.

\section{Summary}
\label{sec:conclusions}
In summary, much progress has been made in the quest for locating a possible QCD critical point. On the theory side many calculations using different methods seem to converge on a region which should be accessible by heavy-ion experiments at collision energies around $\sNN \simeq 5 \gev$.
On the experimental side, the STAR collaboration has delivered excellent, high-statistics data for proton number cumulants from the very successful second phase of the RHIC beam energy scan. At the same time several non-critical baselines have been established which agree with measurements for collision energies above $\sNN \gtrsim 10 \gev$. In our view this makes it very unlikely that a QCD critical point is located at values of the baryon number chemical potential below $\mu_B\lesssim 400 \mev$. However, for the lowest two energies of the beam energy scan, at 9 and 7.7 GeV, the data for the second- and third-order factorial cumulants show significant non-monotonic behavior which is not reproduced by the baselines. These may indicate the onset of attractive interactions and thus be a first hint at critical dynamics. However, there may be other, more mundane effects such as volume or impact parameter fluctuations which seem to result in a similar behavior. These become increasingly important as one lowers the collision energy and therefore need to be understood as one explores the energy regime where a possible QCD critical point is predicted to be. 

In any case, we are just at the beginning of understanding the results from the RHIC BES-II, and more data from the fixed-target program are expected to be available soon. Understanding those data quantitatively may already reveal some more intriguing hints for a QCD critical point. If not, at the very least, it will prepare us for the upcoming CBM experiment, which will be able to measure right where the QCD critical point is predicted to be.

\newpage

\backmatter

\bmhead{Acknowledgments}
This material is based upon work supported by the U.S. Department of Energy, Office of Science, Office of Nuclear Physics, under contract numbers DE-AC02-05CH11231 and DE-SC0026065.





\bibliography{bibfile}
\end{document}